\documentclass[aps,pra,twocolumn,floatfix,superscriptaddress]{revtex4}  
\usepackage{float}
\usepackage{graphicx}
\usepackage{amssymb}
\usepackage{dsfont}
\usepackage{amsmath}  
\usepackage{epstopdf}
\begin{document}

\def\ket#1{|#1\rangle} 
\def\bra#1{\langle#1|}
\def\av#1{\langle#1\rangle}
\def\dkp#1{\kappa+i(\Delta+#1)}
\def\dkm#1{\kappa-i(\Delta+#1)}
\def\pp{{\prime\prime}}
\def\ppp{{\prime\prime\prime}}
\def\w{\omega}
\def\wa{{\omega_a}}
\def\wb{{\omega_b}}
\def\k{\kappa}
\def\D{\Delta}
\def\wp{\omega^\prime}
\def\wpp{\omega^{\prime\prime}}

\title[]{Input-output wavepacket description of two photons interacting with \\ a V-type three-level atom in an optical cavity}
\author{Arkan Hassan}
\author{Julio Gea-Banacloche}%
 \email{jgeabana@uark.edu.}
\affiliation{ 
Department of Physics, University of Arkansas, Fayetteville, AR 72701
}%


\date{\today}

\begin{abstract}
We study the interaction of a V-type atom in a cavity with incident single- and two-photon wavepackets and derive an exact formula, valid in all parameter regimes, relating the spectrum of the outgoing wavepackets to the incident one.  We present detailed results for several special input pulses, and consider the potential performance of the system as a CPHASE gate for initial pulses in a product state.  We find values of the cavity, atomic and pulse parameters that yield a conditional phase shift of $\pi$, albeit with a relatively small overlap between the incoming and outgoing pulse forms.

\end{abstract}

\maketitle

\section{\label{sec:level1}Introduction and summary}

The possibility of quantum information processing using photons as qubits was suggested early on \cite{chuang1}, but the difficulty in producing optical nonlinearities strong enough to enable conditional logic between individual photons held the field back until the proposal by Knill, Laflamme and Milburn \cite{klm} to combine linear optics with probabilistic, measurement-based gates, and the later incorporation of cluster-state computing ideas \cite{cluster1,cluster2} (see Ref.~\onlinecite{kok} for an early review of the field).  A large amount of work has been done in this direction since then, and it is generally regarded, at the moment, as probably the most promising approach to photonic quantum computing.

Nevertheless, interest in alternative ways to achieve conditional logic between single photons has persisted.  On the one hand, deterministic schemes involving the interaction of the photons with single atoms in optical cavities have been proposed \cite{duan,koshino} and, in some cases, demonstrated \cite{rempe}.  Crucial to these schemes is that the photons interact sequentially, not simultaneously, with the atom-cavity system that mediates the interaction, and in between interactions some additional external manipulation is typically required.  Alternative schemes involving the simultaneous interaction of photons as ``flying qubits'' with an effective nonlinear medium were shown to potentially suffer from serious limitations \cite{shapiro}, including spectral entanglement of the photons that would drastically reduce their overlap with their initial, pre-interaction state \cite{kerr} (this was also shown to be the case if the nonlinear medium was placed in an optical cavity \cite{bala}).  Several proposals to mitigate or eliminate this problem altogether have been put forward in recent years, however \cite{knight,brod1,brod2,bala2,konyk,sorensen}, typically involving the further propagation of the photons through a specially-designed medium that has the effect of ``cleaning up'' the wavefunction. As an example, in Refs.~\onlinecite{brod1,brod2} this is achieved by enforcing momentum conservation by propagation through a periodic array of scatterers, such as three-level, V-type atoms, in a waveguide.

The potential of V-type three-level atoms to mediate nonlinear interactions between photons has itself long been recognized, as are the difficulties inherent in ensuring a sufficiently strong coupling between the atoms and the photons to carry out a successful conditional quantum logical gate.  While the above-mentioned schemes would rely on one-dimensional waveguides, the possibility of placing the atom in an optical cavity instead was carefully analyzed by Chudzicki, Chuang and Shapiro \cite{chuang}, who also proposed a method involving repeated photon-cavity interactions, with ``repair'' operations being performed on the pulse after each interaction.  These authors worked in the bad-cavity limit, using a result for the output pulse in this limit due to Koshino \cite{koshino2,koshino3}.   


In this paper we revisit the  problem of a single-sided optical cavity, containing a single V-type atom and coupled to the photon field, in order to provide a complete picture of the pulse distortions, valid in every regime from the good to the bad cavity limits.  We will use the space-time description of the interaction developed by one of us in Ref.~\onlinecite{jgb2013}, and illustrated there for the case of a single two-level atom in a cavity interacting with single-photon wavepackets.  We first consider as input the (exponentially rising) wavepacket that achieves perfect excitation of the atom in the single-photon case, and discuss the differences encountered in the two-photon case.  We  then  explore the potential of the system to serve as a conditional phase gate (CPHASE) for a family of time-symmetric input states.  Finally, for completeness, and also to illustrate the connection between our methods and previous results, we consider also the case where the input is an entangled state that is preserved by the interaction, up to an overall phase shift, in the bad cavity limit \cite{koshino2}.

\section{Model and general solution}
\subsection{Formalism}
The situation we consider is illustrated in Figure 1.  Pulses containing one or two photons, described by their (joint) frequency spectrum, are incident on a one-sided optical cavity containing a single V-level-type atom. Each photon is tuned (by its polarization, for instance) to one of the $a$ or $b$ transitions. The cavity decay rate is denoted by $\kappa$, and the atom-cavity coupling constant is $g$.  Our goal is to express the frequency spectrum of the outgoing pulses as a function of the incoming one.

\begin{figure}[h]
    \includegraphics[width=8cm]{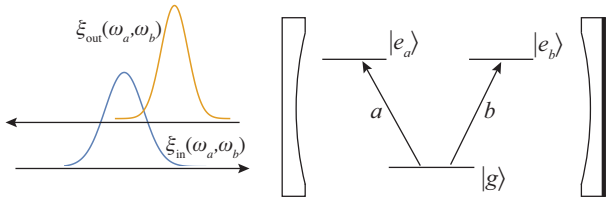}%
    {\caption{Sketch of the experimental situation considered.}}
\label{fig:fig1}
\end{figure} 

We adopt the model of Ref.~\onlinecite{jgb2013}, which was already used to study nonlinear interactions between photons in a cavity in Ref.~\onlinecite{bala}.  The ``modes of the universe'' that describe the (one-dimensional) quantized electromagnetic field both inside and outside the cavity are described by operators $a_{\omega}$ and $b_{\omega}$ that satisfy the continuum commutation relations $[a_\omega,a^\dagger_{\omega^\prime}] = \delta(\omega-\omega^\prime)$,  $[b_\omega,b^\dagger_{\omega^\prime}] = \delta(\omega-\omega^\prime)$.  The frequencies $\omega$ will be understood to be defined relative to the cavity frequency $\omega_c$, that is, $\omega$ is in each case the true optical frequency minus $\omega_c$.  Detuning of the pulses relative to the cavity can always be incorporated in the description of the initial pulse by having it centered at a frequency $\omega$ other than zero.

The field inside the cavity is described by ``single quasimode operators'' $\hat A$ and $\hat B$, which, in the interaction picture, read
\begin{align} 
 \hat{A}(t) &= \int d \omega \frac{\sqrt{\kappa/ \pi}}{\kappa - i \omega}\, {a}_{\omega} e^{-i \omega t}  \cr 
 \hat{B}(t) &= \int d \omega \frac{\sqrt{\kappa/ \pi}}{\kappa - i  \omega}\, {b}_{\omega} e^{-i \omega t}   
\end{align}
The interaction is then described by the Hamiltonian
\begin{equation}
H = \hbar g\left( \hat A(t) e^{-i\delta t}\ket{e_a}\bra g + \hat B(t) e^{-i\delta t}\ket{e_b}\bra g + H.c.\right)
\label{e2}
\end{equation}
where $\delta$ is the detuning between the atomic transition frequency (assumed the same for both transitions) and the cavity frequency: $\delta = \omega_c - \omega_{at}$.  

We assume spontaneous emission is negligible during the interaction time, in which case a pure-state description is adequate.  Then, assuming no more than two excitations are present in the system, the wavefunction at any time $t$ can be written as
\begin{align}
\ket{\Psi(t)} &= \int d\wa \int d\wb \, \xi_{ab}(\omega_a,\omega_b,t) a_\wa^\dagger b_\wb^\dagger \ket{00}\ket{g} \cr
&+\int d\wa \, \xi_a(\wa,t)a_w^\dagger \ket{00}\ket{e_b} \cr
&+ \int d\wb\,  \xi_b(\wb,t)b_w^\dagger \ket{00}\ket{e_a}
\end{align}
where $\ket{00}$ is the field vacuum state.  This yields the Schr\" odinger equations for the spectral functions $\xi_{ab},\xi_a, \xi_b$,
\begin{align}
 \frac{d}{dt} \xi_{ab}(\wa,\wb,t) &=  \frac{-i g \sqrt{\kappa/\pi}}{\kappa+ i \wa} e^{i(\wa+\delta) t} \xi_b(\wb,t) \cr &\qquad - \frac{i g \sqrt{\kappa/\pi}}{\kappa+ i \wb} e^{i(\wb+\delta) t} \xi_a(\wa,t)\cr
 \frac{d}{dt} \xi_{a}(\wa,t) &= \int d\w \frac{-i g \sqrt{\kappa/\pi}}{\kappa - i \w} e^{-i(\w+\delta) t} \xi_{ab}(\wa,\w,t) \cr
\frac{d}{dt} \xi_{b}(\wb,t) &= \int d\w \frac{-i g \sqrt{\kappa/\pi}}{\kappa - i \w} e^{-i(\w+\delta) t} \xi_{ab}(\w,\wb,t)\cr
\label{e4}
\end{align}

\subsection{The single-photon case}
When a single photon is incident on the system, the atom just acts like a two-level atom and the problem reduces to the one solved in Ref.~\onlinecite{jgb2013}.  In this subsection we summarize the most relevant results of that paper, and also show how to derive the basic result using the Laplace transform method that we will adopt later, since there are some subtleties that need to be pointed out.

According to Eq.~(46) of Ref.~\onlinecite{jgb2013}, if a single-photon is incident on the cavity with initial spectrum $\xi(\w,0)$, and with the atom initially in the ground state, the spectrum of the outgoing pulse, which we may denote as $\xi_\text{out}(\w)$, is given by
\begin{equation}
\xi_\text{out}(\w) = -\frac{(\omega-i\kappa)(\omega+\delta)-g^2}{(\omega+i\kappa)(\omega+\delta)-g^2}\,\xi(\omega,0)
\label{e5}
\end{equation}
A special pulse, whose spectrum was denoted by $S^\ast(\omega)$ in Ref.~\onlinecite{jgb2013}), is the maximally coupled one that achieves perfect inversion of the atom at some time, which we will take to be $t=0$ for simplicity.  This spectrum is
\begin{equation}
S^\ast(\omega) = \frac{g\sqrt{\kappa/\pi}}{(\omega-i\kappa)(\omega+\delta)-g^2} 
\label{e6}
\end{equation}
and corresponds to the time-reversed (complex conjugate in the frequency domain) of the pulse emitted by an initially excited atom.  In the time domain, the spectrum (5) describes a rising exponential (with oscillations in the good-cavity, $g>\k/2$ regime; see Fig.~7 of Ref.~\onlinecite{jgb2013})), starting at $t=-\infty$ and ending at $t=0$, whereas the pulse with the conjugate spectrum $S(\omega)$ is a decaying exponential starting at $t=0$.  It is easy to verify that Eq.~(\ref{e5}) transforms an incoming $\xi(\w,0) = S^\ast(\omega)$ into an outgoing $\xi_\text{out}(\w) = S(\omega)$.

We can derive Eq.~(\ref{e5}) from the formalism introduced above as follows.  When only one excitation is initially present (for example, an ``$a$'' photon with the atom in the ground state), the state of the system can be written as
\begin{equation}
\ket{\Psi (t)} = \int d\wa \, \xi_a(\wa,t) a^\dagger_{\wa} \ket{00} \ket{g} + C_{e_a}(t) \ket{00}\ket{e_a} 
\label{e7}
\end{equation}
with the equations of motion
\begin{align}
& \frac{d}{dt} \xi_{a}(\wa,t)= -i g \sqrt{\frac{\kappa}{\pi}} \frac{e^{i(\wa+\delta) t}}{\kappa + i \wa}  C_{e_a}(t) \cr
& \frac{d}{dt} C_{e_a}(t)= -i g \sqrt{\frac{\kappa}{\pi}} \int d\omega \frac{e^{-i(\omega+\delta) t}}{\kappa - i \omega}  \xi_{a}(\omega,t) 
\label{e8}
\end{align}
Introducing the Laplace transforms $\tilde \xi_a(\omega,s)$ and $\tilde C_{e_a}(s)$ of $\xi_a(\omega,t)$ and $C_{e_a}(t)$, we obtain the system
\begin{align}
& s\tilde\xi_{a}(\wa,s)  = \xi_a(\wa,0)-i g \sqrt{\frac{\kappa}{\pi}} \frac{1}{\kappa + i \wa}  \tilde C_{e_a}(s-i(\wa+\delta)) \cr
& s \tilde C_{e_a}(s)= -i g \sqrt{\frac{\kappa}{\pi}} \int d\omega \frac{1}{\kappa - i \omega}  \tilde\xi_{a}(\omega,s+i(\w+\delta)) 
\label{e9}
\end{align}
This is easily solved, by substituting the first equation in the second, solving for $\tilde C_{e_a}(s)$, and substituting back in the first equation.  The result is
\begin{widetext}
\begin{equation}
\tilde \xi_a(\wa,s) = \frac{\xi_a(\w,0)}{s} -\frac{g^2\kappa}{\pi s}\frac{1}{\k+i\wa} \frac{s+\kappa-i\wa}{(s-i(\wa+\delta))(s+\k-i\wa)+g^2}\int\frac{\xi_a(\w,0)}{(\k-i\w)(s+i(\w-\wa))}\,d\w
\label{e10}
\end{equation}
\end{widetext}
This Laplace transform can easily be inverted if one is interested in the time evolution of $\xi_a$ or $C_{e_a}$.  Here we will be concerned only with the long-time result as $t\to\infty$, which is obtained formally as the limit $\xi_a(\wa,\infty) = \lim_{s\to 0} s \tilde\xi_a(\wa,s)$.  Most of the terms in (\ref{e10}) do not pose a problem, but the limit in the integral requires more careful consideration.  

If the incoming pulse vanishes for $t\le 0$, is finite for $t>0$, and vanishes sufficiently fast as $t\to\infty$, its Fourier transform will be an analytic function in the upper half of the complex $\w$ plane and will vanish as $\w\to\infty$ in that region.  The integral can then be evaluated by the residue theorem, over a semicircle in the $\Im(\w) >0$ region, which encloses only the pole at $\w = \wa+is$ and therefore yields $2\pi  \xi_a(\wa+is,0)/(\k-i\wa)$. The limit $s\to 0$ can then be taken without difficulty, and a little further manipulation then yields
\begin{equation}
\xi_a(\w,\infty) = \left(\frac{\w+i\k}{\w-i\k}\right)\frac{(\omega-i\kappa)(\omega+\delta)-g^2}{(\omega+i\kappa)(\omega+\delta)-g^2}\,\xi_a(\omega,0)
\label{e11}
\end{equation}
and the result (\ref{e5}) then follows from noting that, by reasons explained in Ref.~\onlinecite{jgb2013}, the spectrum of the output pulse is related to $\xi(\w,\infty)$ through
\begin{equation}
\xi_\text{out}(\w)= \frac{\kappa+i\w}{\k-i\w}\,\xi(\w,\infty)
\end{equation}
If the incoming pulse starts before $t=0$, but still vanishes for $t<-T$ (with $T>0$), we can formally accommodate this by multiplying its Fourier transform by $e^{i\omega T}$.  This shifts the pulse in time to the right by a distance $T$, so now the function $\xi_a(\w,0) e^{i\omega T}$ satisfies the conditions necessary to integrate over a semicircle in the upper half of the complex $\omega$ plane, and the derivation of (\ref{e11}) will carry through as before, only with $\xi_a(\w,0) e^{i\omega T}$ in place of $\xi_a(\w,0)$.  The phase factor $e^{i\omega T}$ will multiply the final result and can simply be ignored, as it corresponds to just a shift in the origin of time.  

Finally, for a pulse such as (\ref{e6}) that never vanishes for $t<0$, one can still follow this approach, just by making $T$ large enough that the beginning of the pulse is too small to have a physical effect.  In this way, the result (\ref{e5}), which was originally derived by Fourier analysis methods in Ref.~\onlinecite{jgb2013}, may be regarded as universally valid for any input pulse.

In the rest of this paper we will define the fidelity and phase shift jointly as the overlap between the final state of the pulse and some ``ideal'' state, which we expect to be either the input state itself, or related to the input state by some basic transformation.  This overlap can be calculated directly from the spectra of the two pulses:
\begin{equation}
\sqrt{{\cal F}_1}e^{i\phi_1} = \int (\xi_\text{ideal}(\omega))^\ast \xi_\text{out}(\w)\,d\w
\label{e13}
\end{equation}
(the subscript 1 on these quantities refers to the single-photon case).

As mentioned above, the maximally coupled pulse with spectrum (\ref{e6}) produces an identical but time-reversed output, with spectrum $S(\w)$.  These pulses, naturally, do not overlap at all, that is, the inner product $\int (S(\w))^\ast S^\ast(\w)\,d\w =0$.  On the other hand, if we could time-reverse the output pulse, we would have a perfect overlap with the input, since they would be identical.  Put otherwise, if we choose $\xi_\text{ideal}(\w) = \xi(\w,0) = S^\ast(\w)$, we get
\begin{equation}
\sqrt{{\cal F}} e^{i\phi_1} = \int (S^\ast(\omega))^\ast S(\w)\,d\w =0
\label{e14}
\end{equation}
but if we choose $\xi(\w,0) = S^\ast(\w)$ and  $\xi_\text{ideal}(\w) = (\xi_\text{out}(\w))^{TR}$ (where the superscript TR stands for ``time reversed'', i.e., complex conjugation in the frequency domain), we get
\begin{equation}
\sqrt{{\cal F}} e^{i\phi_1} = \int \left(S^\ast(\w)\right)^\ast S^\ast(\w)\,d\w =1
\label{e15}
\end{equation}
  The possibility of time-reversing the output pulse was already mentioned in Ref.~\onlinecite{chuang}.  If it could be done, the choice (\ref{e6}) for the input pulse would be very attractive, since we would not have to worry at all about the pulse distortion introduced by the interaction with the atom-cavity system.  

However, as we shall show in the following Section, the result (\ref{e14}) does not generalize at all to the two-photon case, which forces us to consider alternative input pulse shapes for conditional quantum logic.  The question of how close to 1 the expression (\ref{e13}) can be made for such pulses remains, in a sense, open, depending on the transformations we want to apply to the output pulse.  Nevertheless, for the simple choice $\xi_\text{ideal}(\w) = \xi(\w,0)$, Eq.~(\ref{e5}) suggests that all it should take to make $\cal F \simeq\,$1 is to make the input pulse spectrally narrow enough for the frequency dependence of the phase factor multiplying $\xi(\w,0)$ to be negligible.  We show examples of this for special pulse families in the following Section.

\subsection{The two-photon case}
To solve the system (\ref{e4}) we again perform a Laplace transform and obtain, after considerable manipulation, the following result (see the Appendix for details):
\begin{widetext}
\begin{align} 
&\xi_{ab}(\omega_a, \omega_b, \infty) = \xi_{ab}(\omega_a, \omega_b, 0)-2 g^2\k\left(\frac{1}{K(\wa)P(\wa)}+\frac{1}{K(\wb)P(\wb)} \right)\xi_{ab}(\omega_a, \omega_b, 0) \cr
&\quad +\frac{2g^4\k^2}{\pi}\,\frac{1}{K(\wa)K(\wb)} \left(\frac{K^\ast(\wa)}{P(\wa)}  \lim_{s\to 0}  \int d\wpp  \frac{1}{(s + i (\wpp- \omega_a))}\, \frac{\xi_{ab}(\omega_a + \omega_b -\wpp,\wpp, 0)}{K^\ast(\wpp)P(\wa+ \wb-\wpp)} + \text{same with $a\leftrightarrow b$} \right)\cr
&\quad- \frac{2 g^6 \kappa^2}{\pi}\,\frac{K^\ast(\wa)+K^\ast(\wb)}{K(\wa)K(\wb)}\,\frac{1}{2 g^2 + (\kappa - i (\omega_a + \omega_b + \delta))(2 \kappa - i ( \omega_a + \omega_b))}\left(\frac{1}{P(\wa)} +\frac{1}{P(\wb)}\right) \cr
&\qquad\times\int d\wpp \frac{\xi_{ab}(\omega_a + \omega_b - \wpp,\wpp, 0)}{K(\wpp-\wa-\wb)K^\ast(\wpp) P( \wa+ \wb - \wpp)}
\label{e17}
\end{align}
\end{widetext}
where we have defined
\begin{align}
K(\w) &= \kappa + i\omega \cr
P(\w) &= g^2-i(\w+\delta)(\kappa-i\w)
\end{align}
(Note that $P(\w) = -g\sqrt{\kappa/\pi}/S(\w)$, with $S(\w)$ defined by Eq.~(\ref{e6}).) 

The bad cavity limit, $\kappa \gg g$, can be obtained from the above expression by assuming that $\omega$ is at most of the order of $\gamma \equiv g^2/\kappa \ll\kappa$, in which case one can approximate $P(\w) \simeq \kappa(\gamma -i(\w+\delta))$ and $K(\w) \simeq \kappa$ everywhere.  In this limit the $g^6$ term in (\ref{e17}) is negligible, and we end up with 
\begin{widetext}
\begin{align}
\xi_{ab}(\omega_a, \omega_b, \infty) &= \xi_{ab}(\omega_a, \omega_b, 0)-\left(\frac{2\gamma}{\gamma-i(\wa+\delta)}+\frac{2\gamma}{\gamma-i(\wb+\delta)} \right)\xi_{ab}(\omega_a, \omega_b, 0) \cr 
&\quad +\Biggl[\frac{2\gamma^2/\pi}{\gamma-i(\wb+\delta)} \lim_{s\to 0}\int\frac{d\wpp}{s+i(\wpp-\wb)}\frac{\xi_{ab}(\wa+\wb-\wpp,\wpp)}{\gamma+i(\wpp-\wa-\wb-\delta)} + \text{same with $a\leftrightarrow b$} \Biggr]
\label{e19}
\end{align}
\end{widetext}
As noted in the introduction, this result was derived by Koshino, in the time domain \cite{koshino2,koshino3}, and we have verified the equivalence of the two formulations (see also Section III.C, below).
 
The opposite limit, the ``good cavity'' case where $\kappa \ll g$, is trickier to handle because in that case the atom-cavity spectrum splits into two separate resonances, at $\omega = \pm\sqrt{g^2-\kappa^2/4}$, so we may want to consider pulses centered at $\omega_0 \simeq \pm g$.  In that case, we may want to write, for example, $\w = g+\wp$ around one of those resonances, and expand as follows
\begin{align}
P(\wp) &\simeq -2gi(\k/2-i(\wp\pm\delta/2)) \cr
K(\wp) &\simeq ig
\end{align}
With these substitutions, the last term in (\ref{e17}) is again negligible, and the result becomes formally identical to (\ref{e19}), only with the replacement $\gamma\to \kappa/2$, $\delta\to \delta/2$, and  all the frequencies shifted to the new origin $\omega_0 =g$.

Examples of the evaluation of these expressions for different kinds of pulses are presented in the next section.

\section{spectra and fidelities for selected input pulses}  
\subsection{Full excitation pulse} 
As noted in Section II.B, an exponentially-growing pulse, with the spectrum $S^\ast(\omega)$ given by Eq.~(\ref{e6}), starting at $t=-\infty$ achieves perfect excitation of the atom at $t=0$.  The corresponding output pulse, with spectrum $S(\w)$, vanishes for $t<0$ and is the perfect time-reversed version of the input.  

We can use Eq.~(\ref{e17}) to explore how this works when two photons are incident, that is, when
\begin{equation}
\xi_{ab}(\wa,\wb,0) = S^\ast(\wa)S^\ast(\wb)
\label{e20}
\end{equation}
All the integrals appearing in (\ref{e17}) can be evaluated by the residue theorem; we do not show the explicit expressions here, as they are quite lengthy and yield no particular insights.  

Figure 2 shows the shape (absolute value) of the output pulse's spectrum for $\kappa = 0.335g$, which is in the good cavity regime, so the four characteristic peaks at $\wa, \wb = \pm \sqrt{g^2-\k^2/4}$ are clearly visible.  

\begin{figure}
\includegraphics[width=3.4in]{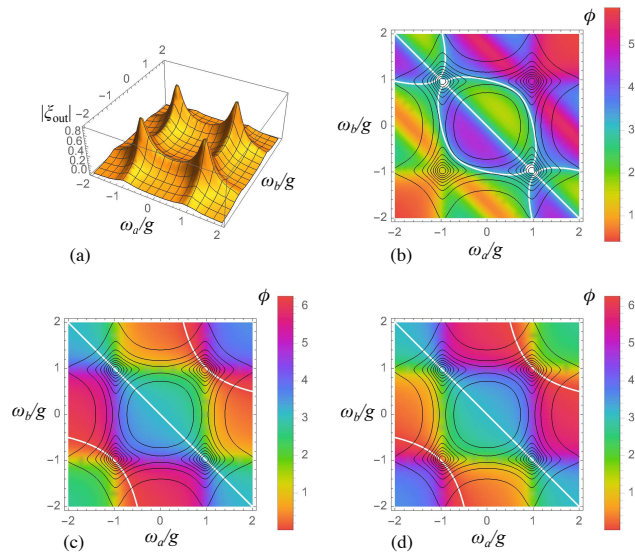}
\caption[example]
   { \label{fig:fig1}
(a) Absolute value of the spectrum of the output pulse, for $\kappa = 0.335g$. (b) Phase plot for the output pulse. (c) Phase plot for the input pulse, and (d) for the time-reversed input pulse, both of these multiplied by $-1$.}
\end{figure}

Interestingly, for the particular choice (\ref{e20}) of input spectrum, we find that the spectrum of the output pulse is equal to just a (frequency-dependent) phase factor times the input spectrum, just as was the case for arbitrary single-photon pulses (see Eq.~(\ref{e5})).  Thus all three pulses---input, output, and time-reversed input---have the same spectral shape.  The phase differences, however, are significant, and determine the final size of their overlap, as calculated by the two-photon generalization of (\ref{e13}):
\begin{equation}
\sqrt{{\cal F}_2}e^{i\phi_2} = \int d\wa \int d\wb \,(\xi_\text{ideal}(\wa,\wb))^\ast \xi_\text{out}(\wa,\wb)
\label{e21}
\end{equation}
Phase maps for all three pulses are also shown in Fig.~2, with contour plots of the spectrum magnitude overlaid on them.  To make the similarities more apparent, the two input spectra have been multiplied by $-1$.  The overlap with the output phase is clearest along the white diagonal line, which corresponds to a phase of $\pi$ in all the plots (the other white lines are curves of zero or $2\pi$ phase). Careful examination shows that the phase overlap is higher for the time-reversed pulse, and indeed, as Figure 3 shows, it is for this choice of $\kappa$ that the overlap of the output pulse with the time-reversed pulse is highest, although still disappointingly far from the perfect overlap that occurred for single-photon pulses.  

\begin{figure}
\includegraphics[width=3.4in]{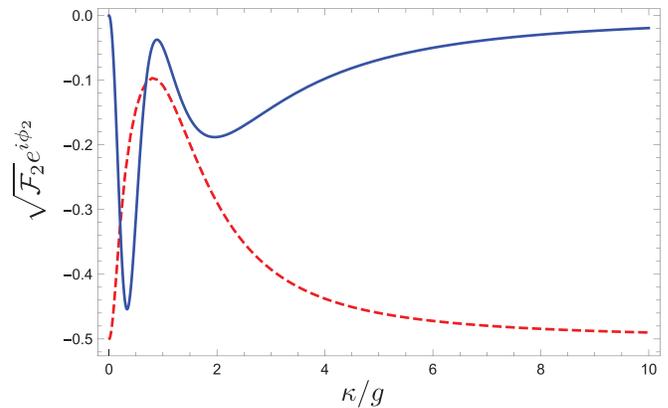}
\caption[example]
   { \label{fig:fig2}
Root of fidelity and phase, $\sqrt{{\cal F}_2}\, e^{i\phi_2}$, for $\xi_\text{ideal}$ equal to the time-reversed input pulse (solid line) and the input pulse itself (dashed line).}
\end{figure}

Another, somewhat surprising, difference with the single-photon case is that now the overlap between the output pulse and the original (not time reversed) input pulse not only doesn't vanish, but can be even larger than for the time-reversed pulse, as seen also in Figure 3 in the extreme limits of a very good ($\kappa/g\to 0$) and very bad ($\kappa/g\to\infty$) cavity.  Nevertheless, again, the highest fidelity obtained is only ${\cal F} = 0.25$, too small to be of any use for quantum information processing (entangling protocols for photons using simple beam-splitters, and just one or two ancillary photons, can more easily achieve success probabilities of this order\cite{linear,linear2}).

\subsection{Lorentzian pulses}

In what follows, we will concentrate on pulses with a Lorentzian spectrum,
\begin{equation}
\tilde f_L(\w,\w_0,\sigma) =\sqrt{\frac{2\sigma^3}{\pi}}\,\frac{1}{(\omega-\omega_0)^2+\sigma^2}
\end{equation}
whose time domain form is 
\begin{equation}
f_L(t) = \sqrt{\sigma}\,e^{-\sigma|t|}
\end{equation}
These pulses capture all the essential features of conventional, time-symmetric pulses (such as Gaussian pulses), and are a convenient choice because all the necessary integrals can be evaluated analytically using the residue theorem.

Our goal is to look for pulses such that the 1 and two-photon fidelities are large, and the difference between the two-photon phase and twice the one-photon phase is also large, ideally close to $\pi$
\begin{equation}
|{\cal F}_1| \simeq |{\cal F}_2| \simeq 1, \quad \phi_2 - 2\phi_1 \simeq \pi
\label{e24}
\end{equation}
These are the conditions one needs to meet in order to perform a conditional phase (CPHASE) gate with high fidelity.  

We will start by looking at what might be the most straightforward way to meet these conditions, namely, to have $\phi_1 = 0$ or $\pi$ and $\phi_2 = \pi$, and also what seems the most natural choice for the pulses' central frequency, namely, $\w_0=\pm\sqrt{g^2-\k^2/4}$ in the good cavity region and $\w_0 =0$ in the bad cavity region, and show the difficulties that arise, before searching numerically for a ``best compromise'' solution.  In all cases, we will use the input pulse as the ``ideal'' output with respect to which we will compute the phase and fidelity, as given by Eqs.~(\ref{e13}) and (\ref{e21}), respectively.

To take the single-photon case first, Eq.~(\ref{e13}), with $\xi_\text{ideal} = \xi(\w,0) = \tilde f_L(\w,\w_0,\sigma)$ yields, using the result (\ref{e5}),
\begin{align}
&\sqrt{{\cal F}_1}e^{i\phi_1} =
\cr
&\quad\frac{(\kappa^2+2\kappa\sigma)(\w_0+i\sigma)^2 + 2 g^2\kappa\sigma +(g^2-(\w_0+i\sigma)^2)^2}
{(\kappa  \sigma +\sigma ^2-\w_0^2-i \w_0
   (\kappa +2 \sigma )+g^2)^2}\cr
   \label{e25}
\end{align}
Figure 4(a) shows the absolute value of this quantity in the good cavity region, whereas Fig.~4(b) shows its actual value in the bad cavity region, since it is a real number for $\w_0=0$. The parameter $r$ in the figures gives the value of $\sigma$ as $\sigma = r \kappa$ and $\sigma = r\gamma$ in the good and bad cavity limits, respectively.  The reason for this scaling is that we do not want our pulse to be much wider than the system's resonance, since in that case most of its spectral components will simply be reflected back, without ever entering the cavity. The plots show that fairly large values of the fidelity are possible for relatively narrow pulses.

\begin{figure}
\includegraphics[width=3.4in]{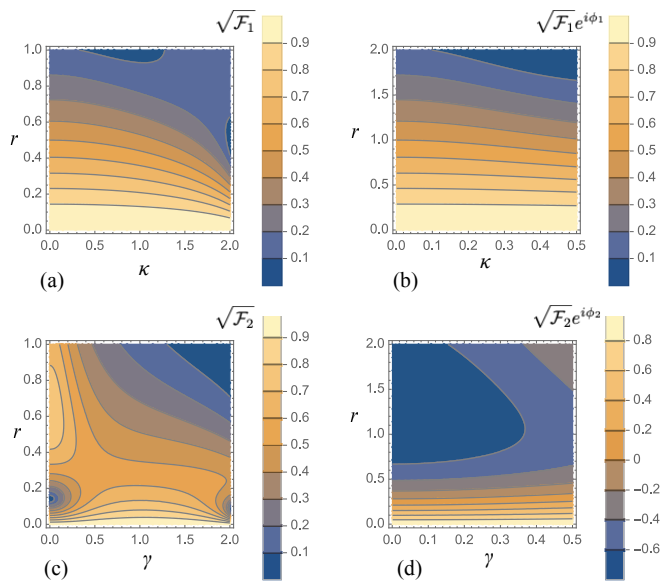}
\caption[example]
   { \label{fig:fig3}
(a) $\sqrt{{\cal F}_1}$ for a single-photon pulse in the good cavity region ($\kappa < 2g$), with $\sigma = r\kappa$ and $\w_0=\sqrt{g^2-\kappa^2/4}$. (b) $\sqrt{{\cal F}_1}e^{i\phi_1}$ (a real quantity here) in the bad cavity region, with $\sigma = r\gamma = r g^2/\kappa$ and $\w_0=0$.  (c) $\sqrt{{\cal F}_2}$ for a two-photon pulse in the good cavity region, parametrized as in (a) above. (d) $\sqrt{{\cal F}_2}e^{i\phi_2}$ in the bad cavity region, parametrized as in (b).}
\end{figure}

The result corresponding to Eqs.~(\ref{e21}) for the two-photon wavepacket can also be calculated analytically, but it is so lengthy that we have preferred not to include it here and discuss it only graphically.  (If desired, it can always, of course, be calculated for specific values of the parameters by numerical integration of the formulas for $\xi_\text{out}$.) Fig.~4(c) shows its magnitude in the good cavity region, and Fig.~4(d) its actual value in the bad cavity region, since, as was the case for the single-photon result, it is actually a real quantity when $\w_0 = 0$. 

The figures show that it is, in general, always possible to get a value of ${\cal F}_i$ close to 1, by making the pulse spectrally narrow enough, which is to say, long enough in time.  When this is done, however, one gets a trivial (and useless for quantum logic) two-photon phase, $\phi_2 = 2\phi_1$.  This is immediately obvious from the right column of Fig.~4, which shows both $\sqrt{{\cal F}_i}e^{i\phi_i}$, $i=1,2$ to be positive quantities for very small $r$.  It is less obvious, but still true, for the small $r$ region of the bad cavity plots (left column of Fig.~4).

The reason for this is simple: When the duration of the pulse is much greater than the lifetime of a photon in the cavity (as modified by the interaction with the atom), the probability that both photons may be simultaneously present in the cavity becomes vanishingly small.  Each photon, therefore, in practice interacts alone with the atom, which for that purpose might as well be a single two-level atom; each individual photon, therefore, merely acquires a phase factor equal to the prefactor of Eq.~(\ref{e5}) evaluated at $\w = \w_0$.

Since the plots in Figure 4 do not really show any other regions with $|{\cal F}_i|$ close to 1, satisfying all the conditions in Eq.~(\ref{e25}) simultaneously appears not to be possible with our choices for $\w_0$.  The best one can do is to use the small-$\gamma$ region in Fig.~4(d) where the two-photon overlap is negative and of the order of $-0.7$.  Unfortunately, in that region ($r>0.7$) the single-photon result is not very large itself.  As an example, for $\kappa/g=20$ ($\gamma = 0.05$) and $r= 0.8$ ($\sigma = 0.04$) we get
\begin{align}  
\sqrt{{\cal F}_1}e^{i\phi_1} &= -0.604 \cr
\sqrt{{\cal F}_2}e^{i\phi_2} &= -0.682 
\end{align}
This gives the desired $\pi$ phase shift, but at the cost of a very low fidelity.

For a most systematic search, it is convenient to define the ``gate fidelity'' in the following way.  Suppose the input  field state is of the form
\begin{equation}
\ket{\Phi_\text{in}} = c_{00} \ket{00} + c_{10} \ket{10} + c_{01} \ket{01} +c_{11} \ket{11} 
\end{equation}
where the $c_{ij}$ are random coefficients whose absolute value squares add up to 1, so $\overline{|c_{ij}|^2} = 1/4$.  The ideal output state would just have the same photonic states except for phases satisfying
\begin{equation}
\ket{\Phi_\text{ideal}} = c_{00} \ket{00} + e^{i\phi_1}c_{10} \ket{10} + e^{i\phi_1}c_{01} \ket{01} - e^{2i\phi_1}c_{11} \ket{11} 
\end{equation}
and hence the gate fidelity ${\cal F}_\text{gate}$ would be, in terms of the quantities we have introduced,
\begin{equation}
{\cal F}_\text{gate} = |\av{\Phi_\text{ideal}|\Phi_\text{out}}|^2 = \frac{1}{16}\left|1 + 2 \sqrt{{\cal F}_1} - \sqrt{{\cal F}_2}e^{i\phi_2-2 i\phi_1} \right|^2
\end{equation}
We have carried out a systematic, computer search for a maximum ${\cal F}_\text{gate}$ over the parameter space, including $\w_0$ a potential detuning $\delta$.  In the good cavity limit, ${\cal F}_\text{gate}$ is maximized for $\w_0 \to \pm g$, $\delta=0$, and $\kappa/g\to 0$ with $\sigma \simeq 0.2742\kappa$.  The asymptotic value of ${\cal F}_\text{gate}$ in this limit is $0.5559\ldots$, or about $9/16$.  To give an idea of how small $\kappa$ needs to be to get close to this value, we may note that, with the above choices, $\kappa = 0.1g$ gives ${\cal F}_\text{gate}\simeq 5.53$. 

Unsurprisingly, given the symmetry between the good and bad cavity limits noted in Section II, we find a corresponding optimal case when $\kappa \gg g$.  This corresponds to $\w_0=0,\delta=0$, and $\kappa/g\to \infty$ with $\sigma \simeq 0.5483\gamma = 0.5483 g^2/\kappa$, and the corresponding asymptotic value is again ${\cal F}_\text{gate}=0.5559\ldots$, exactly the same as in the good cavity case.  Since a bad cavity is much easier to realize than a good one, this would clearly be the choice to go with.  For reference again, we note that $\kappa/g = 25$ already yields ${\cal F}_\text{gate}\simeq 0.555$.

\begin{figure}
\includegraphics[width=3.4in]{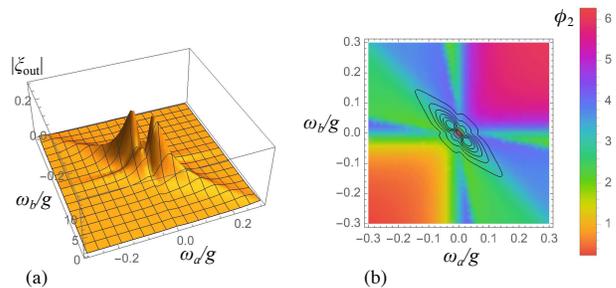}
\caption[example]
   { \label{fig:fig4}
(a) Absolute value of the spectrum of the output pulse, for $\w_0=0$, $\kappa = 25g$, and $\delta=0.0219 g$, giving the optimal gate fidelity in the bad-cavity limit.  (b) Phase plot for the output pulse.}
\end{figure}

The absolute value and phase of the output two-photon pulse's spectrum for this optimal choice of parameters are shown in Figure 5.  Clearly there is considerable distortion of the input pulse (which was originally a single peak) in the absolute value plot, which now shows two peaks along the $\w_a + \w_b=0$ line.  The stretching of the spectrum along this direction is typical of the spectral entanglement one finds in these photon-photon nonlinear interactions\cite{kerr,bala,brod1}.  The phase plot reveals that the ``crater'' at $\w_a=\w_b=0$ serves to suppress the only part of the pulse that does not have a phase shift of approximately $\pi$ radians relative to the input pulse.  

In fact, the two-photon overlap function is real and negative: $\sqrt{{\cal F}_2} e^{i\phi_2} = -0.4824\ldots$, whereas  the single-photon overlap function is real and positive: $\sqrt{{\cal F}_1} e^{i\phi_1} = 0.7489\ldots$, so the condition $\phi_2-2\phi_1=\pi$ is exactly satisfied.  The low gate fidelity results entirely from the small absolute value of the overlap (in both cases, but obviously worse for the two-photon pulse), which measures essentially the probability to detect one or the other (or both) of the output photons in an interference experiment with an unmodified pulse. 


\subsection{Entangled input pulse}
As mentioned in the Introduction, Koshino\cite{koshino2} studied this system in the bad cavity limit, where he addressed the question of what kind of input pulse would actually be left unchanged, except for a constant phase factor of $-1$, by the interaction with the atom-cavity system.  The answer is a two-photon entangled state, and while this is unlikely to be the input state in any quantum computation, we would like to show here, for completeness, how this result follows from our formalism.

Koshino's state would have a temporal profile given by
\begin{equation}
f_K(t_a,t_b) = \frac{1}{\sqrt{2 T_1 T_2}} e^{-|t_a-t_b|/T_1} e^{-|t_a+t_b|/T_2}
\end{equation}
and consequently the corresponding input spectrum would be
\begin{align}
\xi_\text{in}(\wa,\wb) &= \frac{2\sqrt 2}{\pi (T_1 T_2)^{3/2}}\,\frac{1}{(\wa+\wb)^2+1/T_2^2}\cr
&\quad \times\frac{1}{(\wa-\wb)^2+1/T_1^2}
\label{e31}
\end{align}
Substituting (\ref{e31}) in (\ref{e19}) (with $\delta=0$), and carrying out the integration and the $s\to 0$ limit, the output state can be written as the sum of two terms,
\begin{equation}
\xi_\text{out}(\wa,\wb) = \xi_\text{in}(\wa,\wb)(A + B)
\end{equation}
where the $A$ term comes from the first line in (\ref{e19}), and is given by\begin{equation}
A=\frac{i\gamma(\wa+\wb)-\wa\wb-3\gamma^2}{(\gamma-i\wa)(\gamma-i\wb)}
\label{e33}
\end{equation}
and the $B$ term comes from the term in square brackets in (\ref{e19}):
\begin{equation}
B = 4\gamma^2 T_1\frac{2\gamma-i(\wa+\wb)-(\wa-\wb)^2 T_1}{(\gamma-i\wa)(\gamma-i\wb)(1+2\gamma T_1-i(\wa+\wb)T_1)}  
\label{e34}
\end{equation}
In the limit considered by Koshino, where $T_2 \gg T_1 \sim 1/\gamma$, this can be simplified by noting that, by Eq.~(\ref{e31}), $|\wa+\wb| \sim 1/T_2 \ll |\wa-\wb| \sim 1/T_1$.  One can then neglect all the $\wa+\wb$ terms in (\ref{e33}) and (\ref{e34}).  Writing also $\wa\wb = (\wa+\wb)/4-(\wa-\wb)^2/4 \simeq -(\wa-\wb)^2/4$, we get
\begin{equation}
A+B=\frac{-4\gamma^2(3-2\gamma T_1)+(\wa-\wb)^2(1+2\gamma T_1-16\gamma^2 T_1^2)}{(1+2\gamma T_1)(4\gamma^2+(\wa-\wb)^2)}
\end{equation}
which, as noted in Ref.~\onlinecite{koshino2}, becomes simply equal to $-1$ for the special choice (in our notation) $T_1 = -2/\gamma$.

\section{Conclusions}
We have presented a formal solution to the question of how a two-photon pulse is modified by interacting with an optical cavity containing a single V-type three-level atom.  Our result is valid for any value of the ratio $\kappa/g$ and explicitly shows the symmetry between the very good and very bad cavity limits.  

We have studied the effect of the interaction for two special types of pulses.  The first type is the pulse that, in the single-photon case, achieves complete excitation of the atom.  In the single-photon case, the output pulse is exactly the time-reversed of the input pulse.  We found that this symmetry does not hold for an input product state of two photons, each in one of these ``optimal excitation'' wavepackets.  However, the output spectrum is still special in that it equals the input spectrum multiplied by a frequency-dependent phase factor.  This is always the case for single-photon pulses, but not for two-photon pulses, except (as far as we know) for this particular choice of input pulse.

We have then considered a family of input Lorentzian pulses as more typical of what one might encounter if trying to use this system to perform a CPHASE logical gate.  The formal advantage of these pulses is that all the calculations can be carried out analytically, which makes it easy to search the parameter space for optimal solutions.  We have found a maximum value of the gate fidelity of close to $9/16$, which can be achieved asymptotically in both the good and bad cavity limits, but most easily in the latter.  In this limit, the required $\pi$ phase shift is exactly achieved, but the photon itself may not be detected in an interference experiment with the unperturbed pulse.  In other words, we have a probabilistic gate with a success probability of the order of $0.56$ but a \emph{conditional} fidelity of 1 (conditioned on the actual detection of the photon).  

Nevertheless, the complication of having to put an atom in a cavity means this will probably not be a viable alternative to the simpler probabilistic gates of linear optics quantum computation, especially since, once one chooses to go the cavity route, near-deterministic photonic gates can be accomplished by other methods, involving further manipulation of either the atom or the photons (or both).  We note, however, that the recent proposals of Heuck et al. \cite{heuck1,heuck2} to use dynamical coupling to get the photon pulses in and out of the cavity might possibly be also used in conjunction with the V-level scheme to alleviate the difficulties encountered here; this is something we intend to investigate in the near future.  We also note that the bad cavity limit is essentially equivalent to a one-dimensional waveguide, which can perhaps make a solid-state implementation of the system attractive.

Finally, we have considered the special entangled two-photon input, introduced in Ref.~\onlinecite{koshino2}, that is preserved by the interaction, except for an overall $\pi$ phase shift, in the bad cavity limit.  As a sort of ``fixed point'' of the transformation (\ref{e19}), there is a possibility that repeated interactions of an initially unentangled state with the atom-cavity system might converge to such a state. This might also be worth looking into in the future, as might, in general, the possible applications of entangled states produced by this interaction.  We offer this as a parting thought to the memory of Jon Dowling, much of whose work involved exploring new uses for entanglement across a variety of fields.   

\begin{acknowledgements}
J. G.-B. is grateful to the editors of this special Jon Dowling memorial issue for the invitation to contribute. We acknowledge the MonArk NSF Quantum Foundry supported by the National Science Foundation Q-AMASE-i program under NSF award No. DMR-1906383.

\end{acknowledgements}

\section*{Author Declarations}
\subsection*{Conflict of interest}
The authors have no conflicts to disclose.

\section*{Data Availability Statement}
The data that support the findings of this study are available within the article.

\appendix

\section{General solution for a two-photon pulse}

We start from the equations (\ref{e4}). If we redefine
\begin{align}
&\xi^{'}_{ab}(\omega',\omega'',t)= e^{-i(\omega'+\delta) t} e^{-i(\omega''+\delta) t} \xi_{ab}(\omega',\omega'',t)  \cr
& \xi^{'}_{a}(\omega',t)= e^{-i(\omega'+\delta) t}   \xi_{a}(\omega',t) \cr
& \xi^{'}_{b}(\omega'',t)= e^{-i(\omega''+\delta) t}  \xi_{b}(\omega'',t)
\label{a1}
\end{align}
to remove the explicitly time-dependent factors, we get the system
\begin{widetext}
\begin{align}
& \frac{d}{dt} \xi^{'}_{ab}(\omega',\omega'',t)= -i(\omega'+\omega''+2\delta) \xi^{'}_{ab}(\omega',\omega'',t)- \frac{i g \sqrt{\kappa/\pi}}{\kappa+ i \omega'}  \xi^{'}_b(\omega'',t) - \frac{i g \sqrt{\kappa/\pi}}{\kappa+ i \omega''}  \xi^{'}_a(\omega',t)\cr
& \frac{d}{dt} \xi^{'}_{a}(\omega',t)= -i(\omega'+\delta) \xi^{'}_{a}(\omega',t)- \int d\omega'' \frac{i g \sqrt{\kappa/\pi}}{\kappa - i \omega''}  \xi_{ab}(\omega',\omega'',t) \cr
& \frac{d}{dt} \xi^{'}_{b}(\omega'',t)= -i(\omega''+\delta)\xi^{'}_{b}(\omega'',t)- \int d\omega' \frac{i g \sqrt{\kappa/\pi}}{\kappa - i \omega'}  \xi_{ab}(\omega',\omega'',t)\cr
\label{a2}
\end{align}
Taking the Laplace transform, we obtain 
\begin{align}
\tilde{\xi^{'}}_{ab}(\omega',\omega'',s) &= \frac{{\xi}_{ab}(\omega',\omega'',0)}{s+i(\omega'+\omega''+2\delta)} - \frac{i g \sqrt{\kappa/\pi} \tilde{\xi^{'}}_{b}(\omega'',s)}{(\kappa+ i \omega')(s+i(\omega'+\omega''+2\delta))} - \frac{i g \sqrt{\kappa/\pi} \tilde{\xi^{'}}_{a}(\omega',s)}{(\kappa+ i \omega'')(s+i(\omega'+\omega''+2\delta))} \cr
\tilde{\xi^{'}}_{a}(\omega',s) &= \frac{-i g \sqrt{\kappa/\pi}}{s+i(\omega'+\delta)} \int d\omega'' \frac{\tilde{\xi^{'}}_{ab}(\omega',\omega'',s)}{\kappa - i \omega''} \cr 
\tilde{\xi^{'}}_{b}(\omega'',s) &= \frac{-i g \sqrt{\kappa/\pi}}{s+i(\omega''+\delta)} \int d\omega' \frac{\tilde{\xi^{'}}_{ab}(\omega',\omega'',s)}{\kappa - i \omega'}  
\label{a3} 
\end{align}
with the initial condition $ \xi^{'}_{a}(\omega',0)=  \xi^{'}_{b}(\omega'',0) = 0$.  

We solve the system (\ref{a3}) by iteration.  Substituting the first equation into the other two, we obtain
\begin{align}
\tilde{\xi^{'}}_{a}(\omega',s) &= A_{0}(\omega')-\frac{ g^2 \kappa}{\pi} \frac{(s+\kappa+i(\omega'+2\delta))}{\Bigl[\big(s+i(\omega'+\delta)\big) \bigl( s+\kappa +i(\omega'+2\delta)\bigr)+g^2 \Bigr] \big(\kappa+i \omega'\big)} \int  \frac{\tilde{\xi^{'}}_{b}(\omega'',s) d\omega''}{\bigl(\kappa - i \omega''\bigr)\bigl[s+i(\omega'+\omega''+2\delta) \bigr]} \cr
\tilde{\xi^{'}}_{b}(\omega'',s) &= B_{0}(\omega'')-\frac{ g^2 \kappa}{\pi} \frac{(s+\kappa+i(\omega''+2\delta))}{\Bigl[\big(s+i(\omega''+\delta)\big) \bigl( s+\kappa +i(\omega''+2\delta)\bigr)+g^2 \Bigr] \big(\kappa+i \omega''\big)} \int  \frac{\tilde{\xi^{'}}_{a}(\omega',s) d\omega'}{\bigl(\kappa - i \omega'\bigr)\bigl[s+i(\omega'+\omega''+2\delta) \bigr]} 
\label{a4}  
\end{align}
where
\begin{equation}
A_{0}(\omega') = \frac{-i g \sqrt{\kappa/\pi} (s+\kappa+i(\omega'+2\delta))}{\big(s+i(\omega'+\delta)\big) \bigl( s+\kappa +i(\omega'+2\delta)\bigr)+g^2} \int d\omega'' \frac{\xi_{ab}(\omega',\omega'',0)}{(\kappa - i \omega'')\bigl[s+i(\omega'+\omega''+2\delta)\bigr]}
\label{a5}  
\end{equation}
and
\begin{equation}
B_{0}(\omega'') = \frac{-i g \sqrt{\kappa/\pi} (s+\kappa+i(\omega''+2\delta))}{\big(s+i(\omega''+\delta)\big) \bigl( s+\kappa +i(\omega''+2\delta)\bigr)+g^2} \int d\omega' \frac{\xi_{ab}(\omega',\omega'',0)}{(\kappa - i \omega')\bigl[s+i(\omega'+\omega''+2\delta)\bigr]}
\label{a6}  
\end{equation}
depend only on the initial pulse spectrum $\xi_{ab}(\wa,\wb,0)$.

The equations (\ref{a4}) are already written in a way that shows that it is possible to substitute each into the other one, and carry out the integration over either $\wpp$ (in the top equation) or $\wp$ (in the bottom equation), to obtain separate, closed equations for $\tilde{\xi^{'}}_{a}(\wp,s)$ and $\tilde{\xi^{'}}_{b}(\wpp,s)$.  These are 
\begin{align}
\tilde{\xi^{'}}_{a}(\omega',s) & = A_{0}(\omega')-B'_{0}(\omega') +\frac{ g^4 \kappa}{\pi} \frac{(s+\kappa+i(\omega'+2\delta))}{\Bigl[\big(s+i(\omega'+\delta)\big) \bigl( s+\kappa +i(\omega'+2\delta)\bigr)+g^2 \Bigr] \big(\kappa+i \omega'\big) \bigl[(s+\kappa+i \delta)(s+2\kappa+2 i \delta)+g^2 \bigr]} \cr
& \quad \times\int  \frac{\tilde{\xi^{'}}_{a}(\omega''',s) d\omega'''}{\bigl(\kappa - i \omega'''\bigr)\bigl[s+\kappa+ i(\omega'''+2\delta) \bigr]} \cr
\tilde{\xi^{'}}_{b}(\omega'',s) & = B_{0}(\omega'') - A'_{0}(\omega'') +\frac{ g^4 \kappa}{\pi} \frac{(s+\kappa+i(\wpp+2\delta))}{\Bigl[\big(s+i(\omega''+\delta)\big) \bigl( s+\kappa +i(\omega''+2\delta)\bigr)+g^2 \Bigr] \big(\kappa+i \omega''\big) \bigl[(s+\kappa+i \delta)(s+2\kappa+2 i \delta)+g^2 \bigr]} \cr
&\quad \times \int  \frac{\tilde{\xi^{'}}_{b}(\omega'''',s) d\omega''''}{\bigl(\kappa - i \omega''''\bigr)\bigl[s+\kappa+ i(\omega''''+2\delta) \bigr]} 
\label{a7}  
\end{align}
where
\begin{equation}
A'_{0}(\omega'') = \frac{ g^2 \kappa}{\pi} \frac{(s+\kappa+i(\omega''+2\delta))}{\Bigl[\big(s+i(\omega''+\delta)\big) \bigl( s+\kappa +i(\omega''+2\delta)\bigr)+g^2\Bigr] (\kappa+i \omega'')} \int d\omega' \frac{A_{0}(\omega')}{(\kappa - i \omega')\bigl[s+i(\omega'+\omega''+2\delta)\bigr]}
\label{a8}  
\end{equation}
and 
\begin{equation}
B'_{0}(\omega') = \frac{ g^2 \kappa}{\pi} \frac{(s+\kappa+i(\omega'+2\delta))}{\Bigl[\big(s+i(\omega'+\delta)\big) \bigl( s+\kappa +i(\omega'+2\delta)\bigr)+g^2\Bigr] (\kappa+i \omega')} \int d\omega'' \frac{B_{0}(\omega'')}{(\kappa - i \omega'')\bigl[s+i(\omega'+\omega''+2\delta)\bigr]}
\label{a9}  
\end{equation}
again depend only on the initial pulse spectrum, via Eqs.~(\ref{a5}) and (\ref{a6}).
Note that, not only are the equations (\ref{a7}) decoupled, but the integrals that appear in them no longer depend on the independent variables $\wp$ and $\wpp$, so they are just functions of $\k$, $s$ and $\delta$, to be determined.  

To do so, starting with the first equation, define $H_a(\omega,s)= \tilde{\xi^{'}_a}(\omega,s)/(\kappa-i \omega)$ and rewrite the equation in terms of $H_a(\omega,s)$, then divide both sides  by ${s+\kappa+i(\omega'+2\delta)}$ and integrate over $\omega'$.  We obtain
\begin{equation}
\int d\omega' \frac{H_{a}(\omega',s)}{s+\kappa+i(\omega'+2\delta)} = A''_{0}- B''_{0}
\label{a10}  
\end{equation}
where 
\begin{equation}
A''_{0} = \frac{1}{1-(D(s,\k,\delta,g))^2}\int d\omega' \frac{A_{0}(\omega')}{(\kappa-i \omega')\bigl( s+\kappa+i(\omega'+2\delta)\bigr)}
\label{a11}  
\end{equation}
and
\begin{equation}
B''_{0} =  \frac{1}{1-(D(s,\k,\delta,g))^2}\int d\omega' \frac{B'_{0}(\omega')}{(\kappa-i \omega')\bigl( s+\kappa+i(\omega'+2\delta)\bigr)}
\label{a12}  
\end{equation}
and the quantity
$$D(s,\k,\delta,g) \equiv (s+\kappa+i\delta)(s+2\kappa+2 i \delta)+g^2$$
has been defined for convenience.  Note that $A''_{0}$ and $B''_{0}$
no longer depend on any frequency variable.  A similar procedure can be followed for the second equation, resulting finally in 
\begin{align}
\tilde{\xi^{'}}_{a}(\omega',s) & = A_{0}(\omega')-B'_{0}(\omega') +\frac{ g^4 \kappa}{\pi} \frac{\bigl(s+2\kappa+2 i\delta\bigr)\Bigl(A''_{0}-B''_{0}\Bigr)}{\Bigl[\big(s+i(\omega'+\delta)\big) \bigl( s+\kappa +i(\omega'+2\delta)\bigr)+g^2 \Bigr] \big(\kappa+i \omega'\big) D(s,\k,\delta,g)} \cr
\tilde{\xi^{'}}_{b}(\omega'',s) & = B_{0}(\omega'')-A'_{0}(\omega'') +\frac{ g^4 \kappa}{\pi} \frac{\bigl(s+2\kappa+2 i\delta\bigr)\Bigl(B'''_{0}-A'''_{0}\Bigr)}{\Bigl[\big(s+i(\omega''+\delta)\big) \bigl( s+\kappa +i(\omega''+2\delta)\bigr)+g^2 \Bigr] \big(\kappa+i \omega''\big) D(s,\k,\delta,g)} 
\label{a13}  
\end{align}
with
\begin{equation}
B'''_{0} =  \frac{1}{1-(D(s,\k,\delta,g))^2}\int d\omega'' \frac{B_{0}(\omega'')}{(\kappa-i \omega'')\bigl( s+\kappa+i(\omega''+2\delta)\bigr)}
\label{a14}  
\end{equation}
and
\begin{equation}
A'''_{0} =  \frac{1}{1-(D(s,\k,\delta,g))^2}\int d\omega'' \frac{A'_{0}(\omega'')}{(\kappa-i \omega'')\bigl( s+\kappa+i(\omega''+2\delta)\bigr)}
\label{a15}  
\end{equation}
\end{widetext}
The result (\ref{e17}) in the main text can then be reached through a laborious, but ultimately straightforward, process of simplification of the expressions (\ref{a8}), (\ref{a9}), (\ref{a11}), (\ref{a12}), (\ref{a14}) and (\ref{a15}), substituting the result (\ref{a13}) in the equation (\ref{a3}) for 
$\tilde{\xi^{'}}_{ab}(\omega',\omega'',s)$, and taking the long-time limit.  Note that, because of the change of dependent variables in Eqs.~(\ref{a1}), the Laplace transforms of the primed and unprimed variables are related by $\tilde \xi_a(\wa,s) = \tilde \xi^\prime_a(\wa,s-\wa-\delta)$, $\tilde \xi_b(\wb,s) = \tilde \xi^\prime_b(\wb,s-\wb-\delta)$, and $\tilde \xi_{ab}(\wa,\wb,s) = \tilde \xi^\prime_{ab}(\wa,s-\wa-\wb-2\delta)$, so the substitution $s\to s-\wa-\wb-2\delta$ should be performed in $\tilde \xi^\prime_{ab}(\wa,\wb,s)$ before taking the long-time limit $\lim_{s\to 0} (s\tilde \xi^\prime_{ab}$).  Note also that, for simplicity, in deriving Eq.~(\ref{e17}) it has been assumed that the initial spectrum is symmetric in the two variables $\wa$ and $\wb$; however, no such assumption has been made in the derivation of Eq.~(\ref{a13}) above.

\end{document}